\documentstyle[11pt]{article}
\title{\bf Barbero's Hamiltonian derived from a generalized
  Hilbert-Palatini action}
\author{S\"oren Holst\thanks{Department of Physics,
    Stockholm University, Box 6730, S-113 85 Stockholm, Sweden.\newline
    e-mail: holst@vanosf.physto.se}}
\date{6 November 1995}
\begin{document}
\maketitle
\begin{abstract}
Barbero recently suggested a modification of Ashtekar's choice of
canonical variables for general relativity. Although leading to a more
complicated Hamiltonian constraint this modified version, in which the
configuration variable still is a connection, has the
advantage of being real. In this article we
derive Barbero's Hamiltonian formulation from an action, which can be
considered as a generalization of the ordinary Hilbert-Palatini action.
\end{abstract}

In 1986 Ashtekar presented a new pair of canonical variables for the
phase spase of general relativity [1]. These variables led to a much
simpler Hamiltonian constraint than that in the ADM formulation [2],
but had the drawback of introducing complex variables in the phase space
action---something that leads to difficulties with reality conditions
which then must be imposed. A couple of years later the Lagrangian
density corresponding to Ashtekar's Hamiltonian was given
independently by Samuel, and by Jacobson and Smolin [3]. That was seen
simply to be the Hilbert-Palatini (HP) Lagrangian with the curvature
tensor replaced by its self-dual part only.

Recently Barbero pointed out that it is possible to choose a pair of
canonical variables that is closely related to Ashtekars but this time
real [4]. The price paid is that the simplicity of Ashtekars
Hamiltonian constraint is destroyed. However, some advantages are still
present with Barbero's choice of variables. For example, they provide
a real theory of gravity with a connection as configuration variable,
and with the usual Gauss and vector constraint,
thus fitting into the class of diffeomorphism invariant theories
considered in [5] in the context of quantization. In this paper we
derive Barbero's result from an action, and since his formulation
includes also that of ADM and Ashtekar via a parameter, the Lagrangian
density used as starting point in this paper, also includes these
cases. Hence we have found, in a sense, a generalized HP
action.

We now write down this action, and thereafter we will motivate that it
is a good canditate for an action leading to Barbero's formulation,
which then will be explicitly derived from it:
\begin{equation}
  \label{generalizedHP}
  S = \frac{1}{2} \int e e^{\alpha}_{\ I} e^{\beta}_{\ J}
    (F_{\alpha\beta}^{\ \ \ IJ} - \alpha \ast F_{\alpha\beta}^{\ \ \ IJ})
    \equiv \frac{1}{2} \int e e^{\alpha}_{\ I} e^{\beta}_{\ J}
    (F_{\alpha\beta}^{\ \ \ IJ} - \frac{\alpha}{2} \epsilon^{IJ}_{\;\;\;KL}
    F_{\alpha\beta}^{\ \ \ KL})
\end{equation}
Here $e_{\alpha I}$ is the tetrad, $e$ its determinant,
$F_{\alpha\beta}^{\ \ \ IJ}$ the curvature considered as a function of
the connection $A_{\alpha I J}$, and $\alpha$ a (complex) parameter
which will allow us to account for all the cases mentioned above. The
star ($*$) denotes, as is seen, the usual duality operator.

Note that if $\alpha = 0$ (\ref{generalizedHP}) simply is the
HP action, which, when 3+1 decomposed, leads to
the usual ADM formulation. On the other hand, when $\alpha = i$ the
integrand is $e e^{\alpha}_{\ I} e^{\beta}_{\ J} \mbox{}^+ \!
F_{\alpha\beta}^{\ \ \ IJ}$, where $\mbox{}^+ \! F_{\alpha\beta}^{\ \ \
  IJ} (A) = F_{\alpha\beta}^{\ \ \ IJ} (\mbox{}^+ \! A)$ denotes the
self-dual part of the curvature, thus yielding Ashtekar's formulation.
Our claim here is that $\alpha = 1$ leads to Barbero's Hamiltonian
with {\it his} parameter $\beta = 1$. (However our $\alpha$ is not identical
to his $\beta$, rather $\alpha = \beta^{-1}$, as we will see.)

The notation for indices adopted in this article is as follows.
$\alpha$, $\beta$, $\gamma \ldots$ are used as spacetime indices
whereas $a$, $b$, $c \ldots$ denote spatial components. $t$
denotes the time component. $I$, $J$, $K \ldots$ are used as Lorentz
indices and $i$, $j$, $k \ldots$ as spatial such. The time component
of a Lorentz vector is denoted by $0$.

As a first check let us study the variation of (\ref{generalizedHP})
with respect to $A_\beta^{\ \, IJ}$. Using
\[
\delta F_{\alpha\beta}^{\ \ \ IJ} = 2 {\cal D}_{[\alpha} \delta
A_{\beta]}^{\ \, IJ}
\]
where $\cal D$ denotes the covariant derivative acting on both
spacetime and Lorentz indices, we get
\begin{equation}
  \label{dSmapA}
  \delta S = \frac{1}{2} \int e e^{\alpha}_{\ I} e^{\beta}_{\ J}
    (\delta F_{\alpha\beta}^{\ \ \ IJ} - \frac{\alpha}{2}
    \epsilon^{IJ}_{\;\;\;KL}
    \delta F_{\alpha\beta}^{\ \ \ KL}) = \int \delta B_\beta^{\ \, IJ}
    {\cal D}_\alpha (e e^{\alpha}_{\ I} e^{\beta}_{\ J})
\end{equation}
where a partial integration was performed, and where
\[
\delta B_\beta^{\ \, IJ} \equiv \delta A_\beta^{\ \, IJ} - \alpha \ast
\delta A_\beta^{\ \, IJ}
\]
If $\alpha \neq \pm i$ one easily finds $\delta A = \delta A (\delta B)$ from
this, and hence one can choose an arbitrary variation $\delta B$ in
(\ref{dSmapA}). For $\alpha = \pm i$, that is, Ashtekar's case,
$\delta B_\beta^{\ \, IJ}$ clearly is the self-dual (anti self-dual)
part of $\delta A_\beta^{\ \, IJ}$ and the action contains only
$\mbox{}^{\pm} \! F_{\alpha\beta}^{\ \ \ IJ} (A) = F_{\alpha\beta}^{\ \ \
  IJ} (\mbox{}^{\pm} \! A)$. Then we can choose to vary $S$ with respect
to $\mbox{}^{\pm} \! A_\beta^{\ \, IJ}$ instead. So in either case
(\ref{dSmapA}) implies
\[
{\cal D}_\alpha (e e^{\alpha}_{\ [I} e^{\beta}_{\ J]}) = 0
\]
which gives (see for example [6])
\begin{equation}
  \label{A(e)}
  A_{\alpha I J} = e_{\beta I} \nabla_\alpha e^\beta_{\ J}
\end{equation}
Hence, if (\ref{generalizedHP}) is considered as a first order action
it implies, exactly as the ordinary HP action does, that
$A_{\alpha I J}$ is the Levi-Civita spin-connection.

Before we perform the 3+1 decomposition of (\ref{generalizedHP}) we
convince ourselves that it gives the right theory---that is, general
relativity---for all complex values of $\alpha$, and not only for
$\alpha = 0$ or $\alpha = i$ when it is known to do that. Study the
second term in the action:
\begin{equation}
  \label{alfaterm=0}
  e e^{\alpha}_{\ I} e^{\beta}_{\ J} \epsilon^{IJKL} F_{\alpha\beta K
    L} = e_{\gamma}^{\ K} e_{\delta}^{\ L} \epsilon^{\alpha \beta
    \gamma \delta} F_{\alpha \beta KL} = \epsilon^{\alpha \beta
    \gamma \delta} R_{\alpha \beta \gamma \delta}
\end{equation}
Here $R_{\alpha \beta \gamma \delta}$ is the Riemann tensor, and
(\ref{A(e)}) was used in the last step, meaning that this only is true
when the evolution equations is used. But this is clearly equal to
zero, since $R_{[\alpha \beta \gamma]}^{\ \ \ \ \ \, \delta} = 0$.

Thus (\ref{generalizedHP}) differ from the case $\alpha = 0$, that is
ADM, by at most a canonical transformation, so we are indeed working
with the right theory. Of course, for this conclusion to be valid it
would suffice if (\ref{alfaterm=0}) was a total derivative---it does
not have to be zero. But in our case it should be,
since the canonical transformation that we are heading at is of the
form
\begin{equation}
\label{cantransf}
\left\{ \begin{array}{l}    A_a^{\; i} \longrightarrow A_a^{\prime\; i} +
    \gamma \frac{\delta f[E]}{\delta E^a_{\; i}} \\
    E^a_{\; i} \longrightarrow  E^{\prime a}_{\;\; i} =  E^a_{\; i}
  \end{array}
\right.
\end{equation}
where $\gamma$ is some parameter, and
\begin{equation}
  \label{f[E]}
  f[E] = \int \Gamma_a^{\; i} [E] E^a_{\; i} \hspace{2cm} {\rm with}
  \hspace{5mm} \Gamma_{ai} = -\frac{1}{2} \epsilon_{ijk} e_b^{\;j}
  \nabla_a e^{bk} ,
\end{equation}
$\Gamma$ thus being the (spatial) Levi-Civita spin-connection. Then
the functional derivative of $f[E]$ equals $\Gamma$, and a simple
calculation shows that
\[
\int \dot{A_a^{\prime\; i}} E^{\prime a}_{\; i} = \int \dot{A_a^{\; i}}
E^a_{\; i}
\]
Thus all transformations of the form (\ref{cantransf}) correspond to
no change in the Lagrangian density.

Now that we have collected enough confidence that
(\ref{generalizedHP}) is the action we are looking for, let
us do the 3+1 decomposition to verify that this indeed is the case. We
write the time component of the tetrad as
\begin{equation}
  \label{3+1tetrad}
  e_{0I} = N n_I + N^a e_{aI}
\end{equation}
where $n_I$ is the normalized gradient to the time coordinate function
defined on the spacetime, and hence orthogonal to surfaces with $t =
\mbox{\sc Constant}$. More precisely $n^I e_{aI} = 0$ and $n^I n_I = -1$.
$N$ and $N^a$ are the usual lapse and shift, respectively. Now we
choose the so called ``time gauge'': We choose the tetrad in such a
way that $n_I = (1,0,0,0)$. This simply means that the spatial vectors
of the tetrad $e_{\alpha i}$ now span the tangent space to a $t = \mbox{\sc
  Constant}$ surface, and that $e_{a0} = 0$. Of course, this gauge
choice does not put any restrictions on the ADM metric itself.

With the time gauge imposed the 3+1 decomposition of
(\ref{generalizedHP}) looks like
\begin{equation}
  \label{S}
  S = \frac{1}{2} \int \epsilon^{abc} \epsilon^{ijk} e_{ai} (e_{bj}
  \hat{F}_{ctk0}
  + N^d e_{dj} \hat{F}_{bck0} + \frac{1}{2} N \hat{F}_{bcjk})
\end{equation}
where
\begin{equation}
  \label{Fhatt}
  \begin{array}{l}
    \hat{F}_{ctk0} = 2 \partial_{[c} A_{t]k0} + 2 A_{[c|k|}^{\;\;\;\;\;\;m}
    A_{t]m0} - \alpha \epsilon_{kmn} (\partial_{[c} A_{t]}^{\;\;mn} +
    A_{[c}^{\;\;mp} A_{t]p}^{\;\;\;\;n} + A_{[c}^{\;\;m0}
    A_{t]0}^{\;\;\;\;n})
    \\
    \hat{F}_{bck0} = 2 \partial_{[b} A_{c]k0} + 2 A_{[b|k|}^{\;\;\;\;\;\;m}
    A_{c]m0} - \alpha \epsilon_{kmn} (\partial_{[b} A_{c]}^{\;\;mn} +
    A_{[b}^{\;\;mp} A_{c]p}^{\;\;\;\;n} + A_{[b}^{\;\;m0}
    A_{c]0}^{\;\;\;\;n})
    \\
    \hat{F}_{bcjk} = 2 \partial_{[b} A_{c]jk} + 2 A_{[b|j|}^{\;\;\;\;\;\;m}
    A_{c]mk} + 2 A_{[b|j|}^{\;\;\;\;\;\;0} A_{c]0k} - 2 \alpha
    \epsilon_{jkm} (\partial_{[b} A_{c]}^{\;\;m0} +
    A_{[b}^{\;\;mp} A_{c]p}^{\;\;\;\;0})
  \end{array}
\end{equation}
and where $\epsilon^{abc} \equiv \epsilon^{tabc}$, $\epsilon^{ijk}
\equiv \epsilon^{0ijk}$ and $\epsilon_{ijk} \equiv - \epsilon_{0ijk}$

The only terms in (\ref{S}) containing time derivatives are
\[
-E^{ck} \partial_t (A_{ck0} - \frac{\alpha}{2} \epsilon_{kmn} A_c^{\;mn})
\]
where the useful identity
\[
\frac{1}{2} \epsilon^{abc} \epsilon^{ijk} e_{ai} e_{bj} = e e^{ck}
\equiv E^{ck}
\]
was used. This motivates the introduction of new variables
\begin{equation}
  \label{defcalA}
  \left\{
    \begin{array}{l}
      \mbox{}^{\mbox{}^+} \!\! {\cal A}_{ck} \equiv A_{ck0} +
      \frac{\alpha}{2}
      \epsilon_{kmn} A_c^{\;mn} \vspace{2mm} \\
      \mbox{}^{\mbox{}^-} \!\! {\cal A}_{ck} \equiv A_{ck0} -
      \frac{\alpha}{2}
      \epsilon_{kmn} A_c^{\;mn} \\
    \end{array}
  \right.
\end{equation}
with the inverse
\begin{equation}
  \label{invdefcalA}
  \left\{
    \begin{array}{l}
      A_{ck0} = \frac{1}{2} (\mbox{}^{\mbox{}^+} \!\! {\cal A}_{ck} +
      \mbox{}^{\mbox{}^-} \!\!
      {\cal A}_{ck}) \vspace{2mm} \\
      A_c^{\;ij} = \frac{1}{2 \alpha} \epsilon^{kij}
      (\mbox{}^{\mbox{}^+} \!\!
      {\cal A}_{ck} - \mbox{}^{\mbox{}^-} \!\! {\cal A}_{ck})
    \end{array}
  \right.
\end{equation}

Substituting $(A_{ck0}, A_c^{\;mn})$ for $(\mbox{}^{\mbox{}^+} \!\!
{\cal A}_{ck},
\mbox{}^{\mbox{}^-} \!\! {\cal A}_{ck})$ in (\ref{Fhatt}) gives
\begin{eqnarray}
    \hat{F}_{ctk0} & = & - \partial_t \mbox{}^{\mbox{}^-} \!\! {\cal
      A}_{ck} +
    \partial_c \left( A_{tk0} - \frac{\alpha}{2} \epsilon_{kmn}
    A_t^{\;mn} \right)
    +  \epsilon_{klm} A_{t0}^{\;\;\;m} \left( \frac{\alpha^2+1}{2\alpha} \;
    \mbox{}^{\mbox{}^+} \!\! {\cal A}_c^{\;l} +
    \frac{\alpha^2-1}{2\alpha} \;
    \mbox{}^{\mbox{}^-} \!\! {\cal A}_c^{\;l} \right) + \mbox{} \nonumber \\
    & & \mbox{} - A_{tkl} \mbox{}^{\mbox{}^-} \!\! {\cal A}_c^{\;l}
    \vspace{2mm} \nonumber \\
    \hat{F}_{bck0} & = & 2 \partial_{[b} \mbox{}^{\mbox{}^-} \!\! {\cal
      A}_{c]k} -
    \frac{\alpha^2+1}{4\alpha} \epsilon_{klm} \mbox{}^{\mbox{}^+} \!\! {\cal
      A}_{[b}^{\;\;\;l} \mbox{}^{\mbox{}^+} \!\! {\cal A}_{c]}^{\;\;\;m} +
    \frac{-\alpha^2+3}{4\alpha} \epsilon_{klm} \mbox{}^{\mbox{}^-} \!\! {\cal
      A}_{[b}^{\;\;\;l} \mbox{}^{\mbox{}^-} \!\! {\cal
      A}_{c]}^{\;\;\;m} + \mbox{} \nonumber \\
    & & \mbox{} - \frac{\alpha^2+1}{2\alpha} \epsilon_{klm}
    \mbox{}^{\mbox{}^+} \!\! {\cal
      A}_{[b}^{\;\;\;l} \mbox{}^{\mbox{}^-} \!\! {\cal A}_{c]}^{\;\;\;m}
    \vspace{2mm} \nonumber \\
    \hat{F}_{bc}^{\;\;\;jk} & = & \epsilon^{ljk} \partial_{[b}
    \left( \frac{\alpha^2+1}{\alpha} \; \mbox{}^{\mbox{}^+} \!\! {\cal
        A}_{c]l} +
    \frac{\alpha^2-1}{\alpha} \; \mbox{}^{\mbox{}^-} \!\! {\cal
      A}_{c]l} \right) -
    \frac{\alpha^2+1}{2\alpha^2} \; \mbox{}^{\mbox{}^+} \!\! {\cal
      A}_{[b}^{\;\;\;j} \mbox{}^{\mbox{}^+} \!\! {\cal A}_{c]}^{\;\;\;k} +
    \frac{3\alpha^2-1}{2\alpha^2} \; \mbox{}^{\mbox{}^-} \!\! {\cal
      A}_{[b}^{\;\;\;j} \mbox{}^{\mbox{}^-} \!\! {\cal A}_{c]}^{\;\;\;k} +
    \mbox{} \nonumber \\
    & & \mbox{} + \frac{\alpha^2+1}{\alpha^2} \; \mbox{}^{\mbox{}^+}
    \!\! {\cal A}_{[b}^{\;\;\;[j} \mbox{}^{\mbox{}^-} \!\! {\cal
      A}_{c]}^{\;\;\;k]} \nonumber
\end{eqnarray}
{}From this we see that $\mbox{}^{\mbox{}^-} \!\! {\cal A}_{ai}$ is the
dynamical variable and $E^{ai}$ its conjugated momentum. Non-dynamical
variables, apart from the lapse $N$ and the skift $N^d$, are
$A_{ti0}$, $A_{tij}$ and $\mbox{}^{\mbox{}^+} \!\! {\cal A}_{ai}$.
Note that if $\alpha = \pm i$ all terms involving $\mbox{}^{\mbox{}^+}
\!\! {\cal A}_{ai}$ disappears, and the constraint below implied by
variation of $\mbox{}^{\mbox{}^+} \!\! {\cal A}_{ai}$ does not exist.
This simplifies things a lot, and, in fact, Ashtekar's Hamiltonian
follows almost immediately. However, for general $\alpha$ we get the
following constraints when varying the Lagrangian density with respect
to $A_{ti0}$, $A_{tij}$ and $\mbox{}^{\mbox{}^+} \!\! {\cal A}_{ai}$
respectively:
\begin{eqnarray}
  \frac{\partial {\cal L}}{\partial A_{tk0}} & = & - \partial_c E^{ck} -
    \epsilon^{ilk} E^c_{\;i} \left( \frac{\alpha^2+1}{2\alpha} \;
    \mbox{}^{\mbox{}^+} \!\! {\cal A}_{cl} + \frac{\alpha^2-1}{2\alpha}
    \; \mbox{}^{\mbox{}^-} \!\! {\cal A}_{cl} \right) = 0 \nonumber \\
  \frac{\partial {\cal L}}{\partial A_{tmn}} & = & \frac{\alpha}{2}
  \epsilon^{kmn} \partial_c E^c_{\;k} - E^{c[m} \mbox{}^{\mbox{}^-}
  \!\! {\cal
    A}_{c}^{\;\;\;n]} = 0 \nonumber \\
  \frac{\partial {\cal L}}{\partial
    \mbox{}^{\mbox{}^+} \!\! {\cal A}_c^{\;l}} & = &
  \frac{\alpha^2+1}{2\alpha} \epsilon_{klm}
  E^{ck} A_{t0}^{\;\;\;m} +
  \frac{\alpha^2+1}{4\alpha} \epsilon^{abc} \epsilon^{ijk}
  \epsilon_{klm} e_{ai} N^d e_{dj} (\mbox{}^{\mbox{}^+} \!\! {\cal
    A}_{b}^{\;m} + \mbox{}^{\mbox{}^-} \!\! {\cal A}_{b}^{\;m}) + \mbox{}
  \nonumber \\
  & & \mbox{} - \frac{\textstyle \alpha^2+1}{\textstyle 4\alpha}
  \epsilon^{abc} \epsilon^{ijk}
  \epsilon_{ljk} \partial_b (N e_{ai}) + \frac{\alpha^2+1}{4\alpha^2}
  \epsilon^{abc} \epsilon_{ilk} N e_a^{\;i}(\mbox{}^{\mbox{}^+} \!\! {\cal
    A}_{b}^{\;k} - \mbox{}^{\mbox{}^-} \!\! {\cal A}_{b}^{\;k}) = 0
  \nonumber
\end{eqnarray}

A lengthy manipulation of these yields
\begin{equation}
  \label{+A{ai}}
  \mbox{}^{\mbox{}^+} \!\! {\cal A}_{ai} = \mbox{}^{\mbox{}^-} \!\!
  {\cal A}_{ai} - 2 \alpha \Gamma_{ai}
\end{equation}
where $\Gamma_{ai}$ is defined as in (\ref{f[E]}), and they also
determine $A_{ti0} = A_{ti0} (\mbox{}^{\mbox{}^-} \!\! {\cal A},
\Gamma, N, N_a)$ which will show up as (part of) a Lagrange multiplier.
(\ref{+A{ai}}) is naturally
taken as a second class constraint in Dirac's notation, since
trivially we have for the conjugated momenta $\mbox{}^{\mbox{}^+} \!
\Psi_{ai}$ to $\mbox{}^{\mbox{}^+} \!\! {\cal A}_{ai}$ that
$\mbox{}^{\mbox{}^+} \! \Psi_{ai} = 0$, and then $\{
  \mbox{}^{\mbox{}^+} \!\! {\cal A}_{ai} - \mbox{}^{\mbox{}^-} \!\!
  {\cal A}_{ai} + 2 \alpha \Gamma_{ai} \; , \mbox{}^{\mbox{}^+} \!
  \Psi_{ai} \}_{PB} = 1$ using the na\"{\i}ve canonical Poisson
Brackets. Hence (\ref{+A{ai}}) should be inserted into the action.
This gives after some algebraic work
\begin{eqnarray}
  \label{resultat}
  S & = & \int \left[ - \dot{\mbox{}^{\mbox{}^-} \!\! {\cal A}_{ck}}
      E^{ck} + \alpha \Lambda_k (\partial_c E^{ck} + \alpha^{-1}
      \epsilon^{kli}
      \mbox{}^{\mbox{}^-} \!\! {\cal A}_{cl} E^c_{\;i}) \right. + N^d
      \; E^{ck} \tilde{F}_{dck} + \mbox{} \nonumber \\
    &   & \left. \mbox{} + N \; \frac{1}{2}
    \epsilon^{abc} e_{ai} (\alpha \tilde{F}_{bc}^{\;\;\;i} - (1 +
    \alpha^2) R_{bc}^{\;\;\;i}) \right] \\
  {\rm where} & & \mbox{}^{\mbox{}^-} \!\! {\cal A}_{ai} = A_{ai0} -
                  \frac{\alpha}{2} \epsilon_{ijk} A_a^{\;jk} \nonumber \\
              & & \Lambda_k = \alpha A_{tk0} + \frac{1}{2} \epsilon_{kmn}
                   A_t^{\;mn} \nonumber \\
              & & \tilde{F}_{abi} = 2 \partial_{[a} \mbox{}^{\mbox{}^-} \!\!
                {\cal A}_{b]i} + \alpha^{-1} \epsilon_{ijk}
                \mbox{}^{\mbox{}^-} \!\!
                {\cal A}_a^{\;j} \mbox{}^{\mbox{}^-} \!\! {\cal
                  A}_b^{\;k} \nonumber \\
              & & R_{abi} = 2 \partial_{[a} \Gamma_{b]i} +
              \epsilon_{ijk} \Gamma_a^{\;j} \Gamma_b^{\;k} \nonumber \\
              & & {\rm with} \hspace{5mm} \Gamma_{ai} = -\frac{1}{2}
              \epsilon_{ijk} e_b^{\;j} \nabla_a e^{bk} \nonumber
\end{eqnarray}

Note that if one uses the definition (\ref{defcalA}) of
$\mbox{}^{\mbox{}^\pm} \!\! {\cal A}_{cl}$ in (\ref{+A{ai}}) one
obtains
\[
A_{aij} = - \epsilon_{ijk} \Gamma_a^{\;k}
\]
that is, $A_{aij}$ is simply the spatial Levi-Civita spin-connection.
Hence, what we effectively have done is to solve for the rotational
(spatial) part of the connection from the evolution equations and to
reinsert it into the action.

Before we comment on the terms in (\ref{resultat}) we should say
something about how our dynamical variable $\mbox{}^{\mbox{}^-} \!\!
{\cal A}_{ai}$ is connected with the dynamical variables used by
Ashtekar and Barbero. From the fact that $A_{\alpha IJ}$ is the
Levi-Civita spin-connection, that is (\ref{A(e)}), we have
\[
A_{ai0} = e^{\beta}_{\;i} \nabla_a e_{\beta 0} \equiv K_{ai}
\]
which in fact is nothing but the
extrinsic curvature of a $t = \mbox{\sc Constant}$ surface.
Hence, our dynamical variable $\mbox{}^{\mbox{}^-} \!\! {\cal A}_{ai}$
can be written
\[
\mbox{}^{\mbox{}^-} \!\! {\cal A}_{ai} = A_{ai0} - \frac{\alpha}{2}
\epsilon_{ijk} A_a^{\;jk} = K_{ai} + \alpha \Gamma_{ai}
\]
which should be compared with Barbero's dynamical variable
\[
\mbox{}^{\mbox{}^{Bar}} \!\!\! A_{ai} = \Gamma_{ai} + \beta K_{ai}
\]
{}From this we see that
\begin{equation}
  \label{jamfBarbero}
  \alpha = \beta^{-1} \hspace{5mm} {\rm and} \hspace{5mm}
  \mbox{}^{\mbox{}^-} \!\! {\cal A}_{ai} = \beta^{-1} \;
  \mbox{}^{\mbox{}^{Bar}} \!\!\! A_{ai} = \alpha \; \mbox{}^{\mbox{}^{Bar}}
  \!\!\! A_{ai}
\end{equation}
Note that for $\alpha = i$, $\mbox{}^{\mbox{}^-} \!\! {\cal A}_{ai}$ and
$\mbox{}^{\mbox{}^{Bar}} \!\!\! A_{ai}$ equals Ashtekar's dynamical
variable up to a factor:
\[
\left. \mbox{}^{\mbox{}^-} \!\! {\cal A}_{ai}  \right|_{\alpha = i} =
  A_{ai0} - \frac{i}{2} \epsilon_{ijk} A_a^{\;jk} = 2
  \mbox{}^{\mbox{}^+} \!\! A_{ai0} = - \mbox{}^{\mbox{}^{Ash}} \!\!\! A_{ai}
\]
and hence
\[
\left. \mbox{}^{\mbox{}^{Bar}} \!\!\! A_{ai} \right|_{\alpha = i} = -i
  \mbox{}^{\mbox{}^-} \!\! {\cal A}_{ai} = i \;
  \mbox{}^{\mbox{}^{Ash}} \!\!\!
  A_{ai}
\]

With (\ref{jamfBarbero}) in mind we recognize the four terms in
(\ref{resultat}) as the kinetical term ``$\dot{q} p$'', the Gauss law
constraint, the vector constraint and the scalar constraint in
Barbero's formulation (where the scalar constraint is of the form in
formula (18) in [4]). Futhermore, one easily finds the ADM or the
Ashtekar formulation by putting $\alpha = 0$ or $\alpha = i$,
respectively, in (\ref{resultat}).

To summarize, we have shown that the action (\ref{generalizedHP}) with
$\alpha = \beta^{-1}$ is the one corresponding to Barbero's
formulation. This was effectively done by solving for the rotational
part of the spin-connection $A_{\alpha IJ}$ from the evolution
equations, and by inserting this result into the action.
For $\alpha = 0$ the action (\ref{generalizedHP}) is the HP action,
leading to ADM when 3+1 decomposed, and for $\alpha = i$ it is the
self-dual part of the HP action, leading to Ashtekar's
Hamiltonian. Hence it could be looked upon as a nice generalization of
the HP action containing Ashtekar's formulation as a
special case.

Finally, I would like to thank Peter Peld\'an, who actually was the one
that suggested that the action (\ref{generalizedHP}) might lead to
Barbero's Hamiltonian, and Ingemar Bengtsson, who persuaded me to
investigate this suggestion, and also cheered me up during the
sometimes tedious tensor manipulations.

\section*{References}
\newcounter{refnr}
\begin{list}
  {\arabic{refnr}}{\usecounter{refnr}}
  \item Ashtekar, A. (1986) {\it Phys.Rev.Lett.} {\bf 57}, 2244.  \\
    Ashtekar, A. (1987) {\it Phys.Rev. D} {\bf 36}, 1587.
  \item Arnowitt, R., Deser, S., Misner, C.W. (1962) in {\it Gravitation: An
      introduction to Current Research} Ed. L Witten (New York, Wiley)
  \item Samuel, J. (1987) {\it Pramana-J. Phys} {\bf 28} L429. \\
    Jacobson, T., Smolin, L. (1988) {\it Class.Quantum Grav.} {\bf
    5}, 583.
  \item Barbero, J.F. (1995) {\it Phys.Rev. D} {\bf 51}, 5507.
  \item Ashtekar, A., Lewandowski, J., Marolf, D., Mour\~ao, J.,
    Thiemann, T., preprint gr-qc/9504018.
  \item Peld\'an, P. (1994) {\it Class.Quantum Grav.} {\bf
      11}, 1087.
\end{list}

\end{document}